\documentclass{epsconf}
\usepackage{graphicx}
\usepackage{wrapfig}
\usepackage{amsmath,amssymb}
\usepackage{caption}
\usepackage{subcaption}
\title{A statistical analysis approach of plasma dynamics in gyrokinetic simulations of stellarator
turbulence}
\author{A. D. Papadopoulos$^1$, \underline{J. Anderson}$^2$, E-J. Kim$^3$, M. Mavridis$^4$ and H. Isliker$^4$}
\institute{$^1$ School of Electrical and Computer Engineering, National Technical University of
Athens, 157 80, Greece\\
$^2$Department of Space, Earth and Environment, Chalmers University of Technology,
SE-412 96 G\"oteborg, Sweden\\
$^3$ Centre for Fluid \& Complex Systems, Coventry University, Priory St,
Coventry CV1 5FB, UK\\
$^4$ Dept. of Physics, Aristotle University of Thessaloniki, 541 24, Thessaloniki, Greece}
\begin{document}
\maketitle

\section{Abstract}
A geometrical method is used for the analysis of stochastic processes in plasma turbulence.
Distances between thermodynamic states can be computed according the thermodynamic length methodology which allows 
the use of a Riemannian metric on the phase space.
A geometric methodology is suitable in order to understand stochastic processes involved
in e.g. order-disorder transition, where a sudden increase in distance is expected. 
Gyrokinetic simulations of Ion-Temperature-Gradient (ITG) mode driven
turbulence in the core-region of the stellarator W7-X, with realistic quasi-isodynamic
topologies are considered. In gyrokinetic plasma turbulence simulations avalanches, e.g. of heat and
particles, are often found and in this work a novel method for detection is investigated.
This new method combines the Singular Spectrum Analysis algorithm
and Hierarchical Clustering such that the gyrokinetic simulation time series is decomposed into
a part of useful physical information and noise. The informative component of the
time series is used for the calculation of the Hurst exponent, the Information Length
and the Dynamic Time. Based on these measures the physical properties of the time
series is revealed.



\section{The SSA \& HC methods}
The SSA analysis for 1D series is a special case of the 2D SSA method. Therefore only the latter is presented. The Gene-data $Q\left(t, \rho \right)$ forms a matrix of size $N_{x} \times N_{y}$. A window $W$ of size $L_{x} \times L_{y}$ is considered, running inside $Q\left(t, \rho \right)$, that is $L_{x} \leq N_{x},~ L_{y} \leq N_{y}$. In the two dimensional SSA, the trajectory matrix is defined as $\mathbf{X}=\left[X_{1}, X_{2}, \ldots, X_{K_{x}\cdot K_{y}}  \right]$. It is a matrix of size $L_{x} \cdot L_{y}\times K_{x}\cdot K_{y}$ where $K_{x}=N_{w}-L_{w}+1$, $w=x,~y$ and its columns $X_{i}$ are vectorizations (vec) of the $L_{x} \times L_{y}$ submatrices $W$, $X_{k+(l-1)K_{x}}=\mbox{vec}\left( W \right)$, $W=X_{k,l}^{\left(L_{x} \times L_{y}\right)} \equiv \left( Q_{i,j}   \right)_{i=k,j=l}^{L_{x}+k-1, L_{y}+l-1  }$. More compactly $\mathbf{X}=\mathbf{T}_{2D}\left(Q\right)$, where $\mathbf{T}$ is the embedding linear operator. It is assumed that the $x$-axis is oriented to the bottom and the $y$-axis to the right of the $Q$ matrix and the origin is the left upper corner.   

The next step of the SSA analysis consists of the decomposition of the trajectory matrix $\mathbf{X}$. In particular the Singular Value Decomposition is applied on the trajectory matrix leading to the decomposition $\mathbf{X}=\mathbf{X}_{1}+\mathbf{X}_{2},\ldots, \mathbf{X}_{d}$, where $\mathbf{X}_{j}=U_{j}V_{j}^{T}\sqrt{\lambda_{j}}$, $j=1,2,\ldots , d$,  the $j^{\mbox{th}}$ singular vectors and value ($\lambda_{j}$) respectively and $d$ the rank of $\mathbf{X}$. The $d$ values and corresponding matrices $\mathbf{X}_{j}$ are grouped into $m$ sets: $\{1,2,\ldots, d \}=\cup_{i=1}^{m} I_{i}$. The grouping of the indices, $j=1,2,\ldots , d$, is achieved via the agglomerative HC method. In the agglomerative HC method a hierarchy of clusters is produced following a bottom-up approach. In particular each observation (index) forms its own cluster and then observations are merged in an additive manner, moving up in the hierarchy. Each observation corresponds to a particular reconstructed matrix $\tilde{X}_{j}$ related to the $j^{\mbox{th}}$ component $\mathbf{X}_{j}$, of the trajectory matrix $\mathbf{X}$. The observation merging is based on a weighted distance matrix $d_{\mathbf{w}}\equiv 1 - \rho_{\mathbf{w}}$, where $\rho_{\mathbf{w}}\left( i,j \right) \equiv \left(\tilde{X}_{i},\tilde{X}_{j} \right)_{\mathbf{w}}/ \left( \left\|  \tilde{X}_{i} \right\| \left\|  \tilde{X}_{i} \right\| \right), ~\left\|  \tilde{X}_{i} \right\| \equiv \sqrt{\left( \tilde{X}_{i}, \tilde{X}_{i}  \right)}_{\mathbf{w}},~\left(X,Y\right)_{\mathbf{w}} \equiv \sum_{l} w_{l} X^{T}_{l,l}Y_{l,l} $. The distance between observations $i,~j$ is given by  by $\left(d_{\mathbf{w}}\right)_{i,j}$. In the first HC step, pairs of observations $i$, $j$ form a cluster if $i=\mbox{agr}_{i^{*}}\min\left(d_{\mathbf{w}}\right)_{i^{*},j}$ . In the second HC step, since clusters have been formed, there must be a definition of cluster distance. This is provided by the linkage clustering, for which there are many choices, single-linkage, complete-linkage, average-linkage, Ward-linkage, etc. The single-linkage clustering is used here, where clusters $A$, $B$, are at a distance $\min\left( \left(d_{\mathbf{w}}\right)_{i,j}, i \in A, j \in B \right)$. The HC steps continue up-wards clustering until the desirable number of clusters, $m$, are formed (the HC stopping criterion in this work) or until the clusters are too far apart to merge. In this study $m=3$ and we consider the classes (clusters), oscillations, trend and noise. Eventually the noise part is removed. 

The reconstructed matrices $\tilde{X}_{j}$ are the result of antidiagonal averaging \cite{Golyandina} followed by inverse embedding, $\mathbf{T}_{2D}^{-1}$, applied to the $\mathbf{X}_{i}$ matrices.

\section{Evaluation of Information Length, Dynamic Time and Hurst Exponent}
From SSA we obtain time subseries $\tilde{X}_{i}\left( \rho \right)$, $i=1,\ldots, d$ of the various SSA components and it is possible to calculate their dynamic time and information length. The dynamic time $\tau \left( t,\rho \right)$ is a time scale over which the probability of the $\tilde{X}_{i}\left( \rho \right)$ changes on average at time $t$ (denoted as $p\left(\tilde{X}_{i}\left( \rho \right),t \right)$). The probability $p\left( \tilde{X}_{i}\left( \rho \right),t \right)$ is estimated from a subset (window) of $\tilde{X}_{i}\left( \rho \right)$ samples of length $W_{L}$ produced around the time index $t$. In particular, first the samples of $\tilde{X}_{i}\left( \rho \right)$ are interpolated to equally spaced time instances (separated by $dt$), then from the samples of each window $w$ of size $W_{L}$ the $p\left( \tilde{X}_{i}\left( \rho \right),t \right)$ is calculated at the time instant $t$ at the middle of the window's time interval. Then the window $w$ is moved (running window) by one time sample and the $p\left( \tilde{X}_{i}\left( \rho \right),t+dt \right)$ is calculated at the time instant $t+dt$, and so on. The calculation of $p\left(\tilde{X}_{i}\left( \rho \right),t \right)$ is done based on the histogram of $w$-samples. Usually the histogram-produced $p\left(\tilde{X}_{i}\left( \rho \right),t \right)$ is non-smooth and a smoothing Gaussian kernel is applied.  
Then $\tau_{i} \left( t, \rho \right)$ of the $\tilde{X}_{i}\left( \rho \right) $ subsequence, is given by \cite{Anderson2020} :
\begin{equation}
\tau_{i} \left( t,\rho \right)^{2}= 1/\int d\tilde{X}_{i}\left( \rho \right) \frac{1}{p\left( 
\tilde{X}_{i}\left( \rho \right),t \right)}\left(   \frac{\partial  p\left( 
\tilde{X}_{i}\left( \rho \right),t \right) }{\partial t} \right)^{2}. \label{dt}
\end{equation}
From the dynamic time the information length , $L_{i}\left( t,\rho \right)$, can be directly calculated \cite{Anderson2020}:
\begin{equation}
L_{i}\left( t,\rho \right)=\int_{0}^{t} ds \frac{1}{\tau_{i} \left( s,\rho \right)}. \label{IL}
\end{equation}

\begin{wrapfigure}{r}{50mm}\centering
\vspace{0cm} 
 \begin{subfigure}[b]{0.3\textwidth}
\includegraphics[width=40mm]{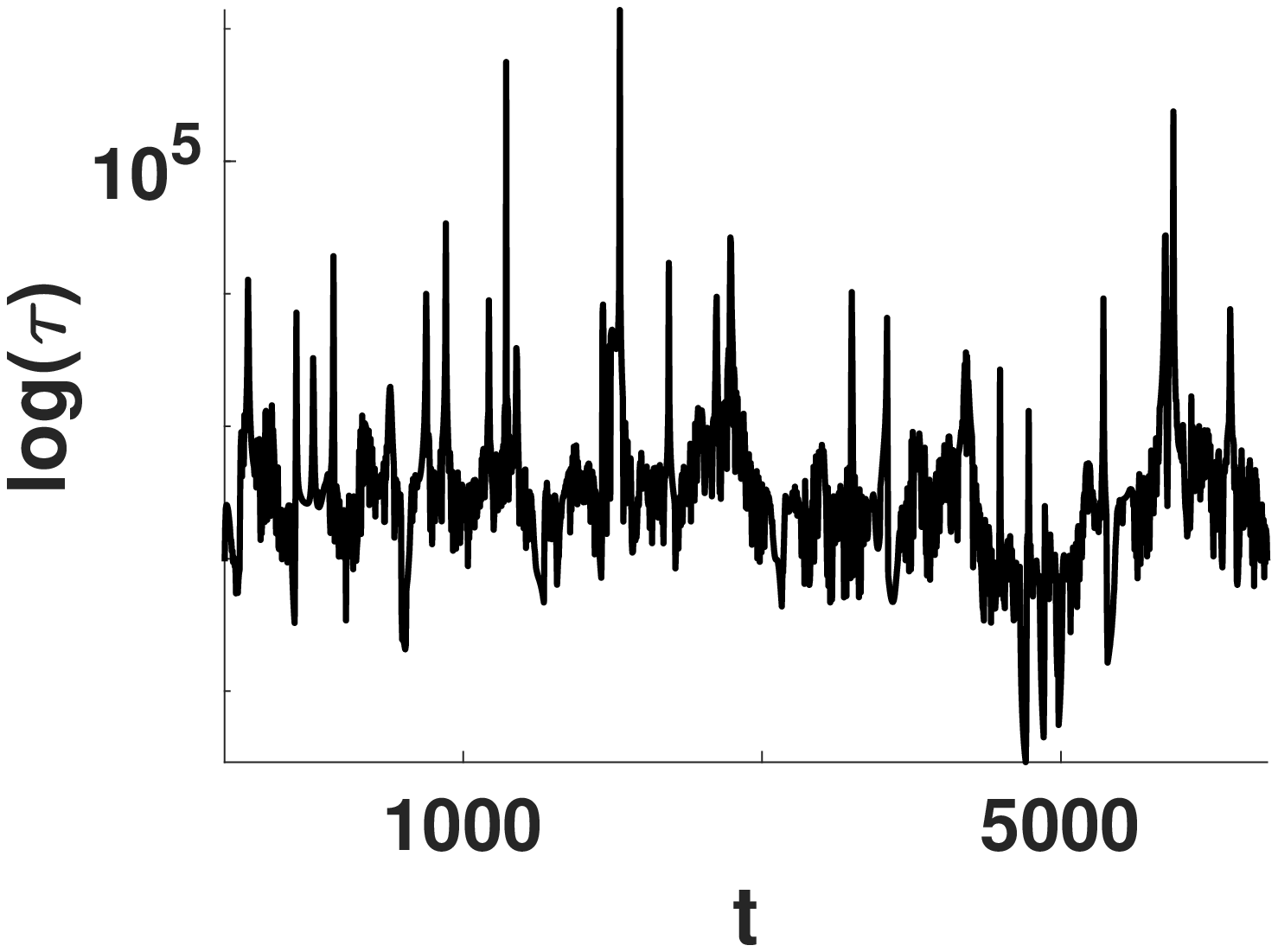}
\caption{\it \small Dynamic time function of time samples $t$.}
\end{subfigure}
\hfill
 \begin{subfigure}[b]{0.3\textwidth}
\centering
\includegraphics[width=40mm]{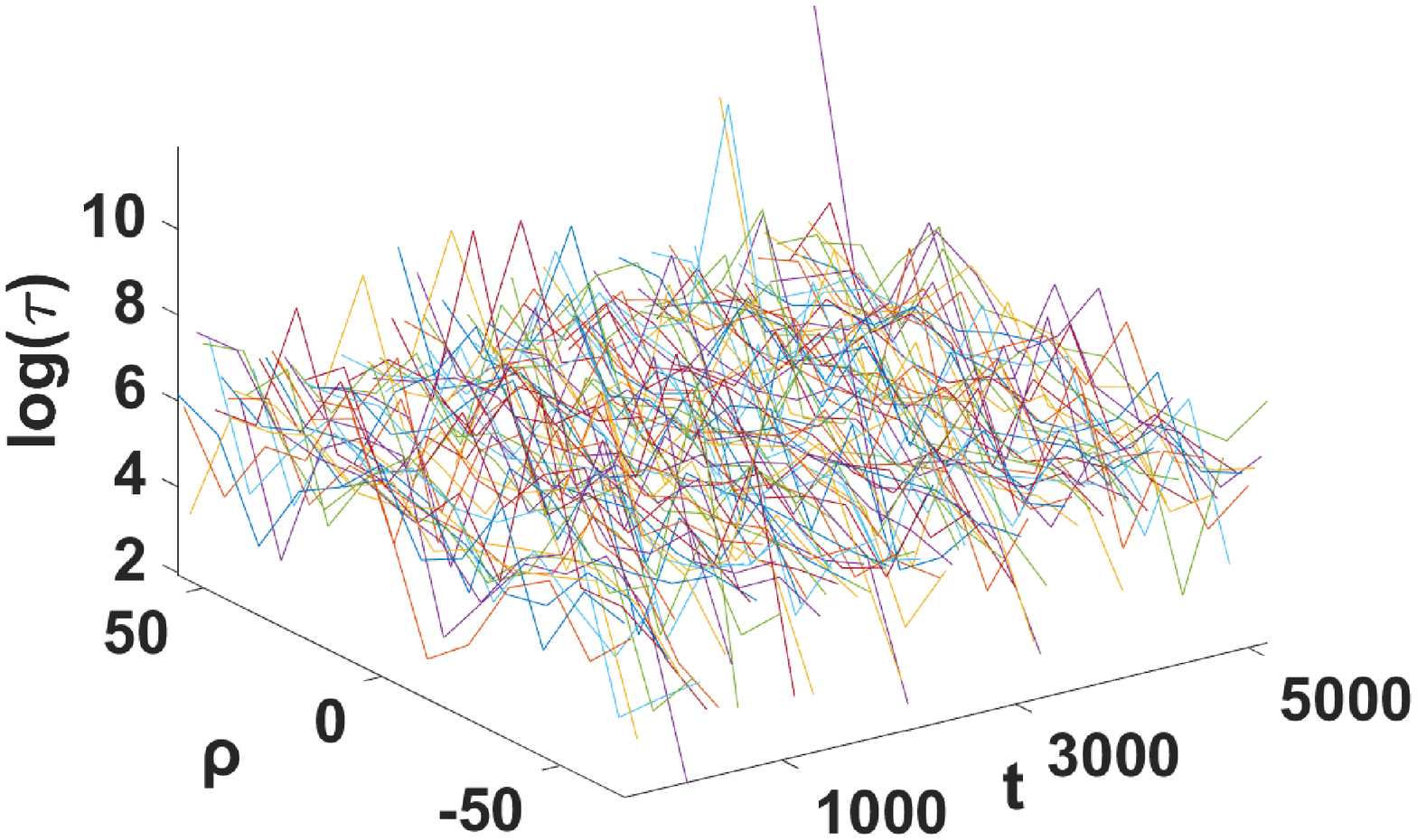}
\caption{\it \small Dynamic time as function of time samples $t$ and varying radius $\rho$.
}
\end{subfigure}
\label{fig1}
\vspace{0cm} 
\end{wrapfigure}

Another quantity of interest is the Hurst exponent, $H$, of $Q\left(\rho \right)$. In general $0 < H < 1$ but if $H = 0.5$ the series is considered random (uncorrelated), 
if $H>0.5$ the series has long term positive autocorrelation meaning that high (low) values in series $Q\left(\rho \right)$ will have a higher probability of being followed by another high (low) value whereas if $H<0.5$ in the long run with high probability high (low) values in  $Q\left(\rho \right)$ will have a higher probability of being followed by 
another low (high) value. The Hurst exponent is calculated by the rescaled range (RS) method as the exponent $H$ such that: $\mathbb{E}\left[ R\left( n \right)/S\left( n \right) \right]=C n^{H}$ for $n \rightarrow \infty$, where $C$ is a constant, $\mathbb{E}\left[\cdot \right]$ is the expected mean, $S\left ( n \right)$ is the standard deviation of the series $Q_{1}\left(\rho \right), Q_{2}\left(\rho \right), \ldots, Q_{n}\left(\rho \right)$ and $R\left ( n  \right)$ is the range of the $n$  cumulative deviations from the mean. That is $R\left( n \right) = \max\left ( Z_{1}, Z_{s}, \ldots, Z_{n}  \right) -\min\left ( Z_{1}, Z_{s}, \ldots, Z_{n}  \right)$, $Z_{j}=\sum _{i=1}^{j}\left( \tilde{X}_{i}-m  \right)$, $m=\left( \sum_{i=0}^{n} \tilde{X}_{i}   \right)/n$. Then $H$ is calculated as the slope of the line that fits the $\log\left ( R(n) / S(n) \right)$ data as  a function of $\log\left ( n\right)$.

\section{Numerical Results} 
The analysis is performed using ion-heat flux data $Q\left(\rho \right)$ time series from the gyrokinetic code GENE. The collected data are for normalized magnetic flux radii $s = 0.5$, normalized temperature gradient $R/L_T =1$ and density gradient $R/L_n =0$. In addition the electrons are considered adiabatic. The $Q\left(\rho \right)$ time series are analyzed with the 1D-SSA  ($Q$ data averaged over the radius $\rho$) or with the 2D-SSA (non averaged $Q$ data). 

The dynamic time stores the instantaneous changes in the information length, this is indicative of sudden changes depending on the present resolution, see Figure 1a and 1b and \cite{Anderson2020}. In Figure 1a the dynamic time in the 1D case is presented, this is average behaviour of the strongly fluctuating signal in Figure 1b. However, the dynamic time exhibit rapid fluctuations and to have a consistent change in dynamics a significant change in the information over a short interval is needed. It is therefore suggested that only by analyzing the information and the dynamic time in tandem, an indication of change in dynamics can be detected. The information length is the additive effect of a series of consistent fluctuations, see Fig 2a and 2b.

\begin{wrapfigure}{r}{50mm}\centering
\vspace{0cm} 
 \begin{subfigure}[b]{0.3\textwidth}
\includegraphics[width=40mm]{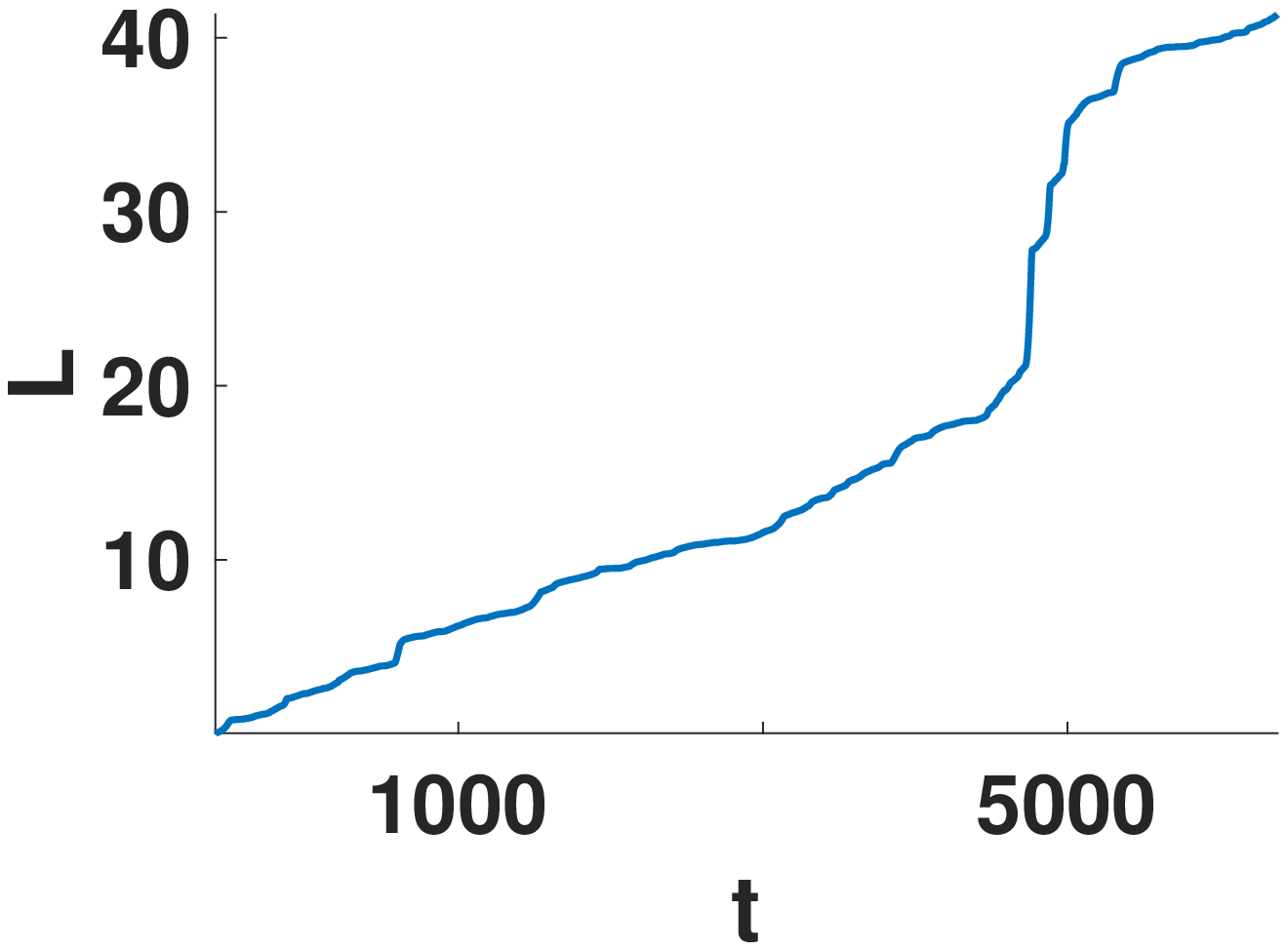}
\caption{\it \small Information length as function of time samples.}
\end{subfigure}
\hfill
 \begin{subfigure}[b]{0.3\textwidth}
\centering
\includegraphics[width=40mm]{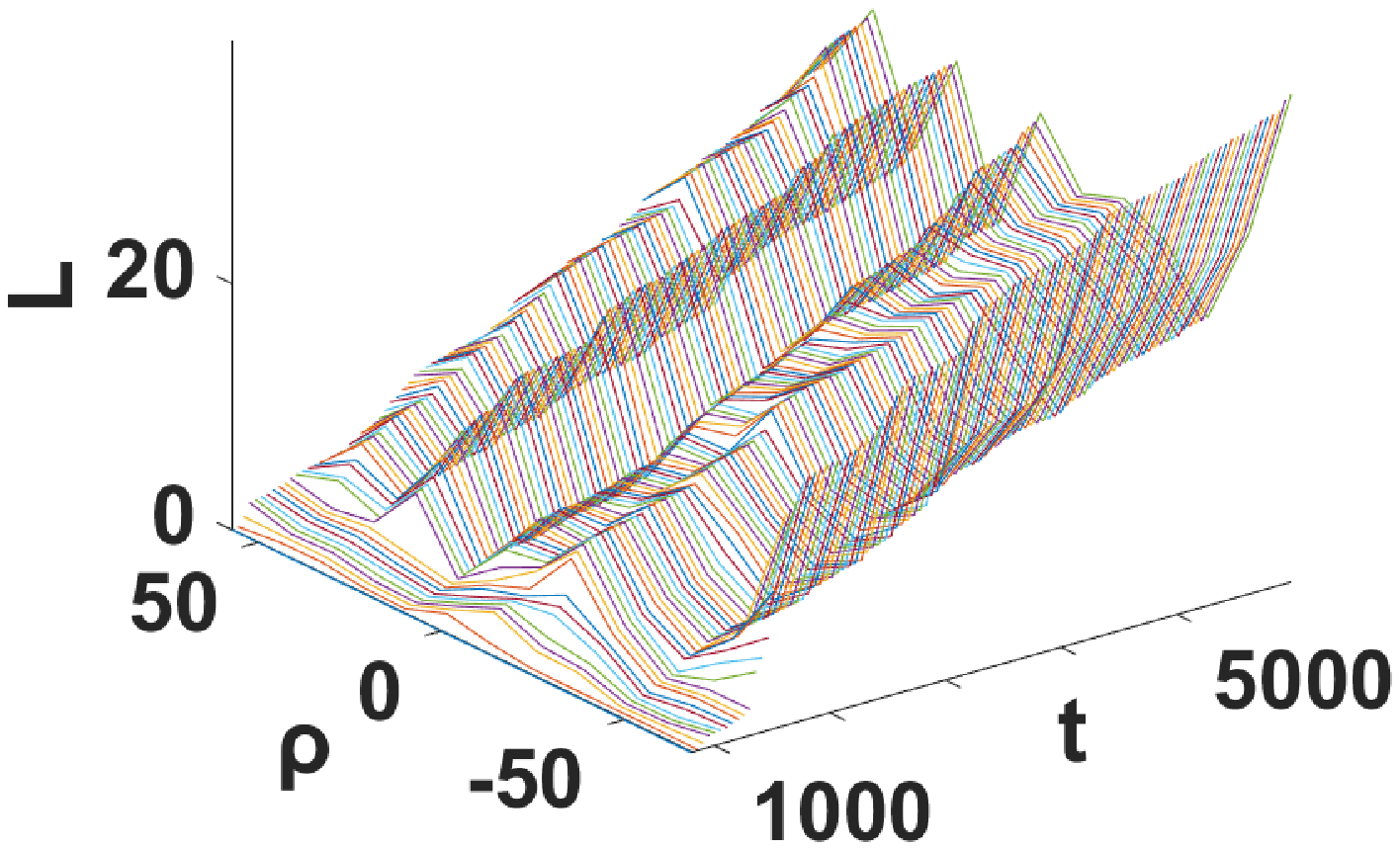}
\caption{\it \small Information length, as function of time samples,
and varying radius $\rho$.   }
\end{subfigure}
\label{fig2}
\vspace{0cm} 
\end{wrapfigure}

A sudden increase in information length is found at around t = 5000 in the 1D case. The information is almost linearly increasing until the change in the dynamics is suggested by a sudden increase in information. In 2D a more varied picture is found. At each radii the information length is calculated, which then gives a almost linear increase in the information length for each radii. It is observed that at certain radii there are sharp increases in information length e.g. $ \rho = 0$ after a certain time similar to the 1D case.

\begin{wrapfigure}{r}{50mm}\centering
\vspace{0cm} 
\includegraphics[width=40mm]{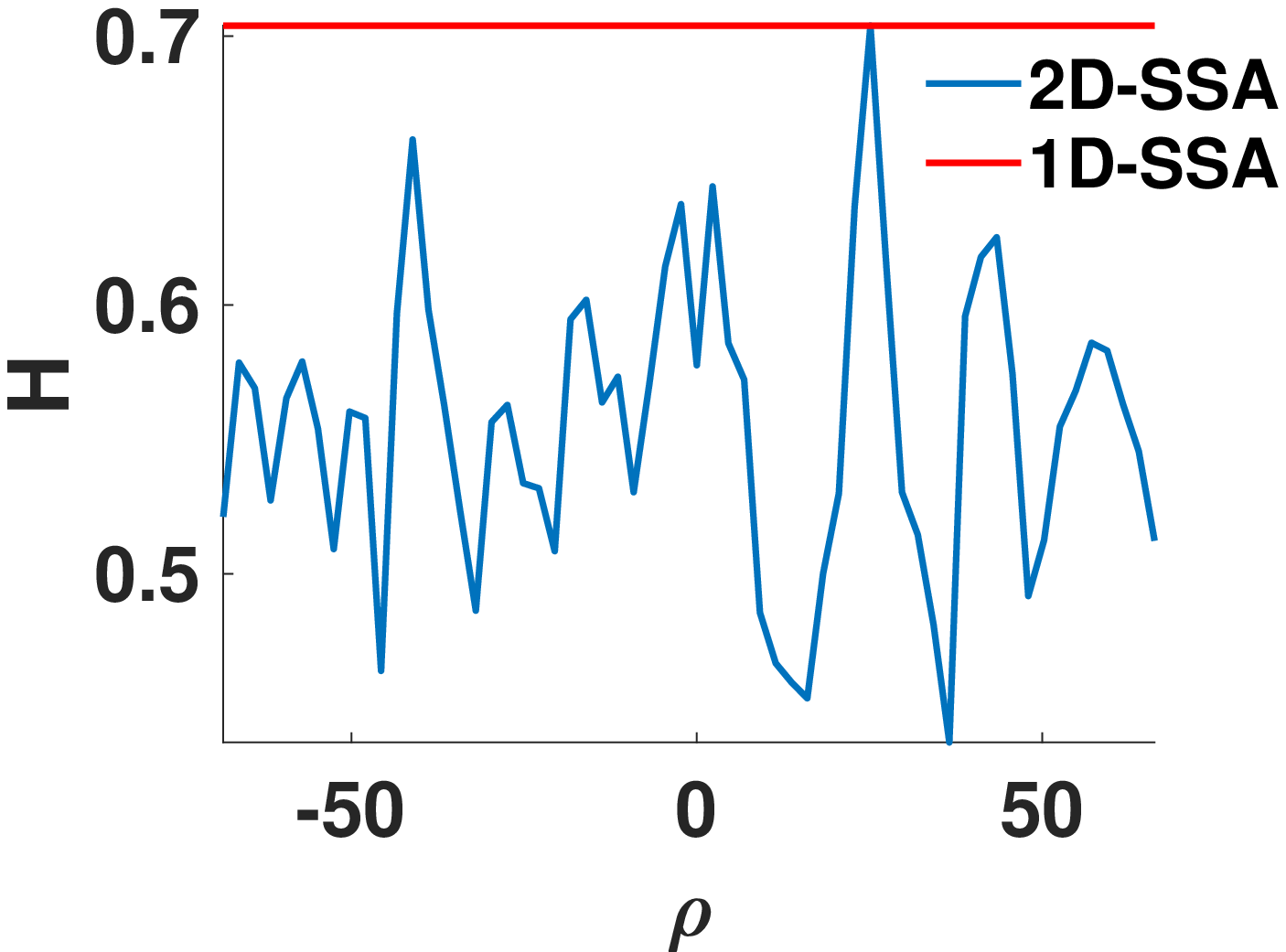}
\caption{\it \small Hurst exponent, $H\left( \rho \right)$ calculated by the 1D and 2D SSA .}
\label{fig3}
\vspace{0cm} 
\end{wrapfigure}

\section{Discussion}

The information length measures the distance between two consecutive states in plasmas. It is thus believed that this measure could be used as indication that some large structure is formed e.g. and is mediating significant transport. In a previous effort the 1D model was developed however it was deemed difficult to directly compare with larger events found in the simulations thus a 2D version is now presented. It is seen that large fluctuations in dynamic time, the instantaneous change in length Fig 1, can be observed if they are consistent c.f. Hurst exponent Fig 3. It is found that at certain radii and times there is a significant increase in information length.

\section{Acknowledgement}

AP was supported by the European Union via the Euratom Research and Training Programme under Grant 101052200 EUROfusion through EUROfusion Consortium.

\end{document}